\documentstyle[preprint,aps]{revtex}

\begin{document}
\draft
\author{A.S. Alexandrov}
\address{Department of Physics, Loughborough University, Loughborough LE11\\
3TU, U.K.}
\title{Comment on `Phase-separated states in antiferromagnetic semiconductors with a
polarizable lattice'}
\maketitle

\begin{abstract}
The 'proof' of the possibility of the slab or stripe phase separation
in polar semiconductors, proposed recently by E. L. Nagaev (Phys. Rev. B${\bf %
64}$ (2001) 014401) is shown to be erroneous. The fundamental error
originates in the double-counting of the electron-phonon interaction
with optical phonons.
\end{abstract}


\narrowtext
 In a recent publication \cite{nag}
Nagaev claims that he  'managed to prove' the possibility  of charge
segregation in antiferromagnetic semiconductors with a polarizable
lattice.
The author neglects the bulk of earlier results including the pioneering
works by Pekar \cite{pek}, Vinetskii and Giterman \cite{vin} and
more recent studies \cite{emi,tak,xom,moi,ita,dev,cat,good,kus,tru,alekab},
where the problem of bound states of a few and many 
 polarons was properly addressed in both large
 and small polaron cases. The
conclusions  of Ref. \cite{nag} are in diametric contradiction with
those and many other studies. Here I show that Ref. \cite{nag} is
wrong on an elementary level.

A 
possibility of pairing of two  large polarons was considered 
by Pekar half a century ago \cite{pek}. He found that a large bipolaron does not exist 
independently of the crystal parameters, $\epsilon_{\infty}$ and $\epsilon_{0}$. Physically, one can reach this 
conclusion by scaling arguments \cite{emi}. The long range interaction with 
optical phonons is  Coulomb like at large distances. 
Then the  total energy of a large polaron, which must be
 minimised is
\begin{equation}
E_{p}(r)={\pi^{2}\hbar^2\over{2mr^{2}}}-{e^{2}\over{2\epsilon^* r}}.
\end{equation}
The first term in Eq.(1) is the kinetic energy of a particle 
confined in a sphere of  radius $r$, while the second term is the 
potential energy  due to lattice polarisation ($1/\epsilon^*=\epsilon_{\infty}^{-1}-\epsilon_{0}^{-1}$). Minimising Eq.(1) with respect to $r$
one obtains the 
polaron binding energy
\begin{equation}
E_{p}= 
-{1\over{4\pi^{2}}}\alpha^{2}\hbar \omega,
\end{equation}
where $\alpha$ is the Fr\"ohlich coupling constant with the optical
 phonons of a frequency $\omega$. For a  
 state of two carriers  
sharing the same orbital within 
a common potential well the 
corresponding functional is
\begin{equation}
E_{b}(r)=2{\pi^{2}\hbar^2\over{2mr^{2}}}-4{e^{2}\over{2\epsilon^* 
r}}+{e^{2}\over{\epsilon_{\infty} r}},
\end{equation}
where the first term is twice the  polaron kinetic energy, the 
second term is four times the corresponding term for a polaron because 
the polarisation is twice as large as that for a polaron, and the last 
term describes the (bare) Coulomb repulsion between two carriers.

The large bipolaron is energetically stable with respect to dissociation 
into two separate large polarons if the binding energy is positive,
\begin{equation}
\Delta \equiv 2E_{p}-E_{b}>0.
\end{equation}
This however is not the case  because \cite{emi}
\begin{equation}
\Delta=2|E_{p}| 
\left[\left(1-{\epsilon_{\infty}\over{\epsilon_{0}-\epsilon_{\infty}}}\right)^{2}-1 \right]<0.
\end{equation}
While a large portion of the Coulomb repulsion is nullified by the 
Fr\"ohlich electron-phonon interaction the latter interaction alone remains 
insufficient to produce a bound state. That is  because 
$\epsilon_{0} > \epsilon_{\infty}$. Approaching the problem from the 
weak coupling limit Takada\cite{tak} and Khomskii \cite{xom} reached the same conclusion.  However, in a strongly 
polarisible lattice with $\epsilon_{0}/\epsilon_{\infty}\gg 1$ the
absolute value of  the bipolaron binding energy 
$|\Delta|$, Eq.(5) 
 is very small 
\begin{equation}
{\Delta \over{2 E_{p}}}\simeq 2{\epsilon_{\infty}\over{\epsilon_{0}}}<<1. 
\end{equation} 
Therefore one can expect that the  wave functions of two polarons  strongly 
overlap, so that the quantum $exchange$ interaction can stabilise 
Pekar's bipolaron even without any other additional  attraction 
but the Fr\"ohlich one.  This was first realised by Vinetskii and
Giterman\cite{vin}, and confirmed in a number of other comprehensive  studies
\cite{moi,ita,dev}. The most reliable path-integral approach
\cite{dev} proved that large bipolarons might be formed if the ratio
of the static and high-frequency dielectric constants is very large,
$\epsilon_{0}/\epsilon{_\infty}\gg 1$. The formation of three-polaron bound states, as well as many-polaronic droplets, slabs,
stripes and strings was shown to be impossible with the long-range Fr\"ohlich or
zero-range Holstein interactions \cite{cat,tru,alekab}. 

Surprisingly, the author of Ref.\cite{nag} reaches the opposite
conclusion. He introduces the Hamiltonian (Eq.(1) of Ref. \cite{nag}) with the Fr\"ohlich interaction,
$H_{sp}$ (Eq.(4) of Ref. \cite{nag}) and the $renormalised$ Coulomb interaction, $H_{c}$,
defined with the $static$ dielectric constant, $H_c \propto
1/\epsilon_0$ (Eq. (11) of Ref. \cite{nag}).  This is  in disagreement with all textbooks, and previous
studies, where the Coulomb interaction is defined with  the
high-frequency dielectric constant, $H_c \propto
1/\epsilon_{\infty}$. As a result, the author finds a 'polaron
instability' occuring at $\epsilon_{0}=2\epsilon_{\infty}$ (Eq.(24) of Ref. \cite{nag}). Analysing the problem of phase-separated states he
further claims that 'if one wishes to take into account the
polaronic contribution, one should replace $\epsilon_{0}^{-1}$ (in the
Coulomb term) by
$\kappa^{-1}=\epsilon_{0}^{-1}-2\epsilon_{\infty}^{-1}$' (Eq.(27) of
Ref. \cite{nag}). This substitution leads him to a conclusion that a charge segregation
appears in manganites with the values of $\epsilon_0=5$ and
$\epsilon_{\infty}=3.4$.
 The author fails to understand that an enhanced value of the
static dielectric constant in polar semiconductors is mainly due to
the Fr\"ohlich interaction itself. By taking the Coulomb repulsion
in his Hamiltonian with the $static$ dielectric constant, he already
accounts for the attractive interaction between carriers mediated
by the optical phonons. Introducing $H_{sp}$ as an independent
term in the Hamiltonian leads to the obvious double-counting of
the electron-phonon interaction with such a choice of the Coulomb
repulsion.  Instead of replacing
$\epsilon_0$ in the incorrect $H_c$ with incorrect $\kappa$, one could
rather  replace the high-frequency dielectric
constant, $\epsilon_{\infty}$, in the $correct$ $H_c \propto
1/\epsilon_{\infty}$ with  $\epsilon_0$ to account for the  Fr\"ohlich
interaction.     Different from 
$\kappa$ in Eq.(27) of  Ref. \cite{nag}, the static dielectric constant is always positive. Hence, the
effective polaron-polaron interaction is always repulsive at large
distancies, so that there is no phase separation of the  Fr\"ohlich polarons
\cite{alekab}. When an additional finite-range 
interaction with the deformation potential or with the
antiferromagnetic fluctuations is  introduced, $E_{att}$, only short-length
stripes are theoretically  possible
\cite{kus,alekab}. The number of polarons bound in the stripe is estimated as \cite{alekab}
\begin{equation}
N = \exp \left(\frac {\epsilon_{0}aE_{att} \delta \omega }{e^{2}\omega} 
-2.31 \right ),
\end{equation}
where $\omega$ is the characteristic frequency 
 of acoustic phonons or  magnons responsible for the finite-range attraction,
$\delta \omega$ its maximum dispersion, and $a$ is the lattice
constant.  Then, with Nagaev's value of $\epsilon_{0}=5$, and  $a=3.8
\AA$,  there is no charge segregation at all
because  $N < 2$ for any  realistic $E_{att} \leq $1 eV.  
Hence,  it is more probable that the polarons in
oxides remain in a charge-honogeneous state. Indeed, there is a growing
understanding that the coexisting phases in manganites must have  nearly the same
charge densities\cite{dag,lit,feh}. Ref. \cite{nag}
is wrong from beginning to end on a rather elementary level. 

I would like to thank Victor Kabanov for calling my attention  to Ref. \cite{nag} and
the EPSRC for partial support (grant R46977).


\begin{references}
\bibitem{nag}
E. L. Nagaev, Phys. Rev. B{\bf 64}, 014401 (2001).
\bibitem{pek}
S. I. Pekar, 'Research in Electron Theory of Crystals', US AEC Report,
AEC-tr 5575 (1963) (Russian original Gostekhizdat (1951)).
\bibitem{vin}
V. L. Vinetskii and M. Sh. Giterman, Zh. Eksp. Teor. Fiz. {\bf 33},
730 (1958) (Sov. Phys.- JETP {\bf 6}, 560 (1958)).
\bibitem{emi}
D. Emin, in 'Polarons and Bipolarons in High-T$_c$ Superconductors and
Related Materials', eds. E. K. H. Salje, A. S. Alexandrov, and
W. Y. Liang, Cambridge University Press (Cambridge, 1995), p.80.
\bibitem{tak}
Y. Takada, Phys. Rev. B {\bf 26}, 1223 (1982).
\bibitem{xom}
D. Khomskii, in 'Polarons and Bipolarons in High-T$_c$ Superconductors and
Related Materials', eds. E. K. H. Salje, A. S. Alexandrov, and
W. Y. Liang, Cambridge University Press (Cambridge, 1995), p.375.
\bibitem{moi}
S. G. Suprun and B. Ya. Moizhes, Fiz. Tverd. Tela (Leningrad) {\bf
  24}, 1571 (1982).
\bibitem{ita}
F. Bassani, M. Geddo, G. Iadonisi, and D. Ninno, Phys. Rev. B {\bf
  43}, 5296 (1991).
\bibitem{dev}
G. Verbist, F. M. Peeters, and J. T. Devreese, Phys. Rev. B {\bf 43},
2712 (1991).
\bibitem{cat}
C.R.A. Catlow, M.S. Islam and X. Zhang, J. Phys.: Condens. Matter 
${\bf 10}$, L49 (1998).
\bibitem{good}
J. S. Zhou and J. B. Goodenough, Phys. Rev. Lett. {\bf 80}, 2665 (1998).
\bibitem{kus}
F.V. Kusmartsev, J. Phys. IV France, {\bf 9}, 321 (1999).
\bibitem{tru}
J. Bonca, T. Katrasnik, and S. A. Trugman, Phys. Rev. Lett. {\bf 84},
3153 (2000). 
\bibitem{alekab}
A. S. Alexandrov and V. V. Kabanov, JETP Lett, {\bf 72}, 569 (2000).
\bibitem{dag}
A. Moreo, M. Mayr, A. Feiguin, S. Yunoki, and E. Dagotto, Phys. Rev. Lett. 
{\bf 84}, 5568 (2000).
\bibitem{lit}
N. D. Mathur and P. B. Littlewood, Solid State Comun. {\bf 119}, 271 (2001). 
\bibitem{feh}
A. Weisse, J. Loos, and H. Fehske, Phys. Rev. B {\bf 64}, 104413 (2001).
\end{references}
\end{document}